\documentclass{article}
\usepackage{frascatiphys,here,graphicx,subfigure}
\begin{document}
\title{CENTRAL PRODUCTION WITH TAGGED FORWARD PROTONS AND THE STAR DETECTOR AT RHIC \\
}
\author{
W\l odek Guryn for the STAR Collaboration        \\
{\em Brookhaven National Laboratory} \\
}
\maketitle
\baselineskip=11.6pt
\begin{abstract}
We describe a setup which will allow extend the physics reach of the STAR detector at RHIC to include the measurement of very forward protons. Tagging on very forward protons, detected by the Roman Pots at RHIC energies, selects processes, in which the proton stays intact and the exchange is dominated by one with the quantum numbers of the vacuum, thus enhancing the probability of measuring reactions where colorless gluonic matter dominates the exchange. The processes include both elastic and inelastic diffraction. The capabilities of the STAR detector to detect Gleuballs and Exotics in central production mechanism are described.
\end{abstract}
\baselineskip=14pt

\section{Introduction}
Installing Roman Pots of the pp2pp experiment ~\cite{pp2ppplb04,lynn,guryn} at STAR ~\cite{star} detector at RHIC will allow tagging events with very forward protons, thus extending physics reach of the experiment to select processes, in which the proton stays intact and the exchange has the quantum numbers of the vacuum. Consequently, enhancing the probability of measuring reactions where colorless gluonic matter dominates the exchange. The processes include both elastic and inelastic diffraction. 

These processes are related to the photon diffraction that has already been studied by 
STAR in Ultra Peripheral Collisions (UPC)of gold-gold (AuAu), and deuteron-gold (dAu) ions, where two pion, $\rho \to \pi^+\pi^-$ and four pion $\pi^+\pi^-\pi^+\pi^-$ photoproduction has been used to probe Pomeron-heavy nucleus couplings~\cite{STARrho,Jsegerphoton2007}. 

\begin{figure}[]
\begin{center}
\includegraphics[width=120 mm]{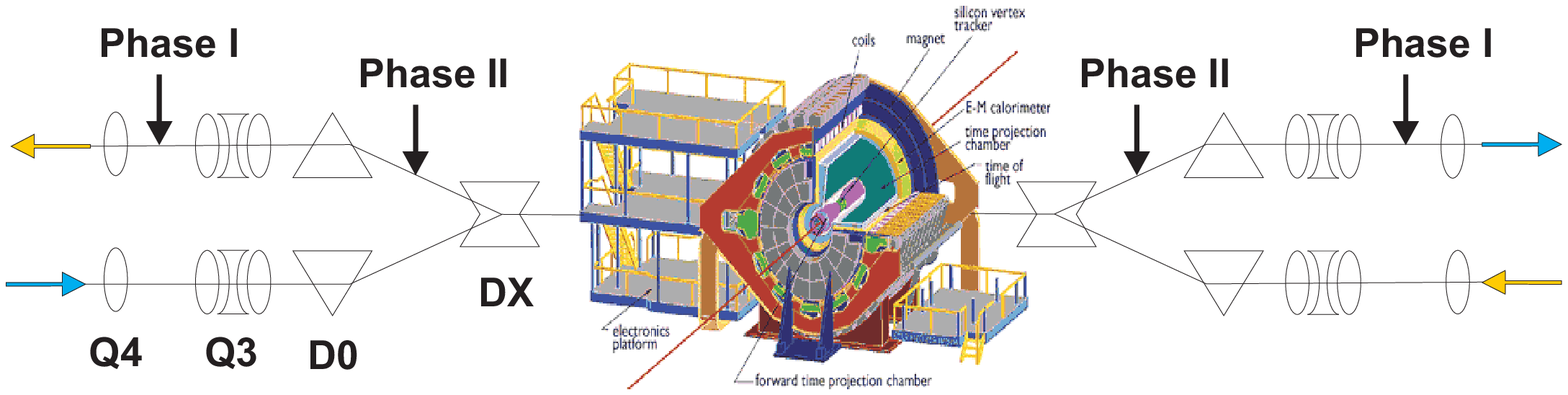}
\caption{\it The Roman pots of the pp2pp experiment in the STAR interaction region, with the arrows indicating proposed locations for Phase I and Phase II.}
\label{layout}
\end{center}
\end{figure}

In order to characterize those diffractive processes well, the measurement of the momentum of the forward proton is important. Because of the layout of STAR and its solenoidal magnetic field, RHIC accelerator magnets must be used for momentum analysis resulting in forward proton taggers installed downstream from the STAR detector, on either side. There are two possible locations: 1) warm section between the DX-D0 magnets; or 2) in the warm straight section between Q3 and Q4 magnets, see fig.~\ref{layout}. Also, to extend the $t$ and $\xi$ ranges to the lowest values, where $t$ is four-momentum transfer between the incoming and outgoing protons, $\xi=\Delta p/p$ is the momentum fraction carried off by the Pomeron,  a moveable detector system, approaching the beam as closely as possible, is needed. Hence the use of pp2pp Roman Pots (RPs) is advantageous not only because it is a working system but because it will also allow maximizing the t and $\xi$ ranges.
We describe a scenario of executing the physics program in two phases, which optimizes the use of available resources and maximizes physics output. Phase I has been implemented and is ready to take data in 2008. In the future the physics reach will be extended to higher values of $t$ and larger data samples will be taken. This would be achieved in Phase II, for which design work is needed.  
\section{Physics Program}
\begin{figure}[]
\begin{center}
\subfigure[]
        {\includegraphics[width=55mm]{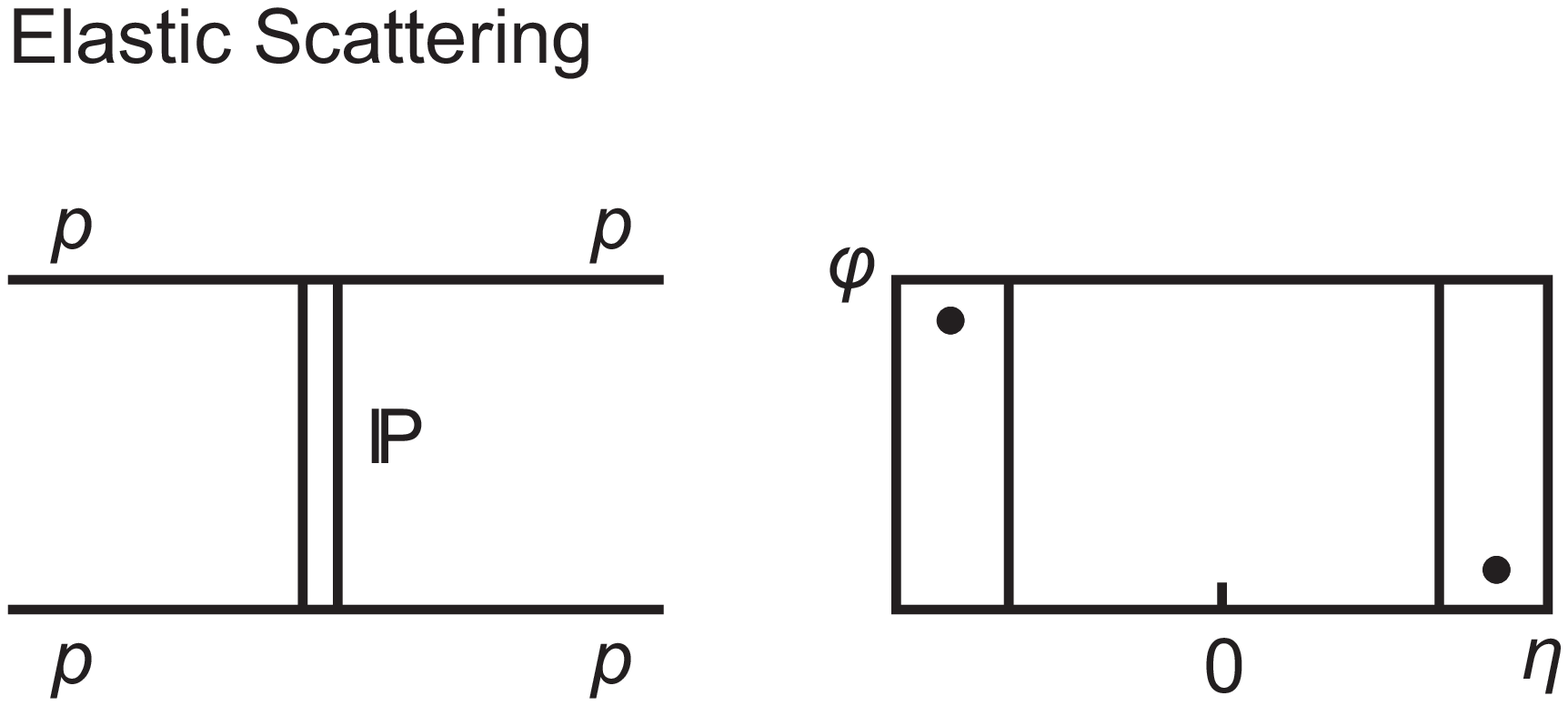}}
\subfigure[]
        {\includegraphics[width=55mm]{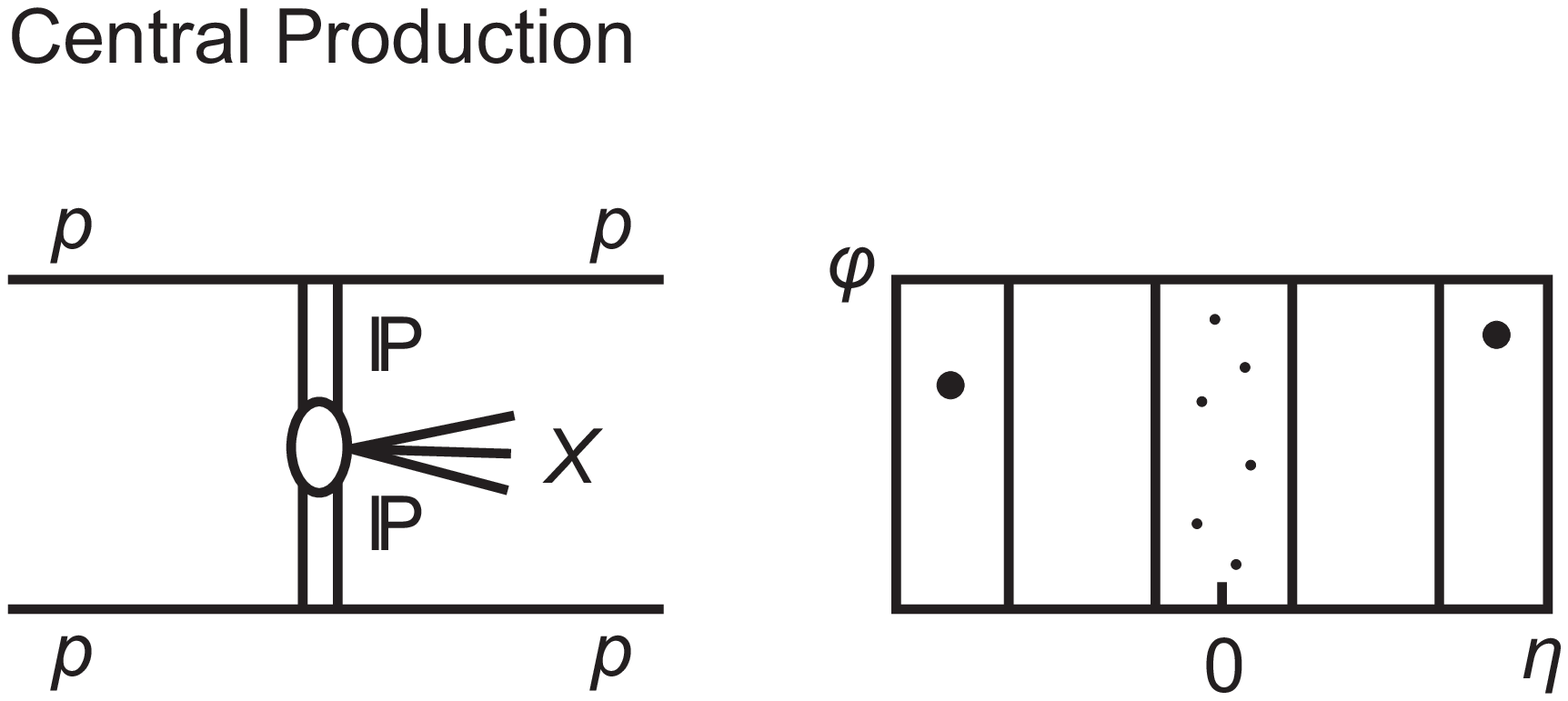}}
        \caption{\it a) Elastic scattering process b) Diffractive - Central Production Process.}
\label{DiffProc}
\end{center}
\end{figure}

In this section we shall briefly describe physics topics in proton-proton collisions, which can be addressed by using the Roman Pots of the pp2pp experiment and the STAR detector. A more detailed description of the physics can be found in~\cite{barrone, donnachie}. At RHIC the processes of interest in polarized proton-proton and proton nucleus collisions are~\cite{bravar}: elastic scattering, central production and single diffraction processes. Here we shall focus on the first two, fig.~\ref{DiffProc}.

The common feature of those reactions is that the proton undergoes quasi-elastic or elastic scattering and that they occur via the exchange of colorless objects with the quantum numbers of the vacuum, historically called Pomeron exchange. In terms of QCD, Pomeron exchange consists of the exchange of a color singlet combination of gluons. Hence, triggering on forward protons at high (RHIC) energies dominantly selects exchanges mediated by gluonic matter. In addition, the use of polarized proton beams, unique at RHIC, will allow exploring unknown spin dependence of diffraction.

Tagging and measuring forward protons also removes the ambiguity of a (complementary) rapidity gap tag, which has a background due to the low multiplicity of diffractive events, and allows the full characterization of the event in terms of $t$, $\xi$ and $M_X$.

\subsection{Elastic scattering}

In studies of the elastic scattering process we will use unique capabilities of RHIC colliding polarized proton beams to measure both spin dependent and spin averaged observables. 

Almost the entire energy range of this proposal has been inaccessible to proton-proton scattering in the past. A measurement of the total cross section, $\sigma_{tot}$ at the highest possible energy will probe the prevalent assumption that the cross sections for $pp$ and $p\bar p$ scattering are asymptotically identical.

The measurement of the differential pp cross section $d\sigma/dt$ over the extended $t$-range will include the region at the lower $|t|$ that is particularly sensitive to the $\rho$-parameter. This will allow extracting the $\rho$-parameter and the nuclear slope parameter $b$ in a combined fit to the differential cross section possible and might also lead to an extraction of $\sigma_{tot}$.

An asymptotic difference between the differential and total cross sections for $pp$ and $p\bar p$ could be explained by a contribution of the Odderon to the scattering amplitude. The absence of an Odderon contribution would lead to identical cross sections, approaching each other roughly as $s^{1/2}$.

By measuring spin related asymmetries one will be able to determine elastic scattering at the amplitude level~\cite{Kopel03,Butt99,Butt01}.  The availability of longitudinal polarization at STAR in this first phase would allow measuring $A_{LL}$ in addition to $A_{NN}$, $A_{SS}$, and $A_{N}$ resulting in a significant improvement of our physics capabilities. Full azimuthal coverage for elastic events has been implemented in this phase.

One of the physics motivations to measure the $A_N$ is to study of the $\sqrt{s}$ dependence of the spin-flip to spin-nonflip amplitudes ratio~\cite{Trueman07}. In other words it may occur that small contribution from hadronic spin-flip to the spin single-spin asymmetry, measured with a polarized jet target at 100 GeV/c, could increase at $\sqrt{s}=200 GeV$. This will help to reveal long standing problem of the energy dependence of the spin flip amplitude, which is best answered experimentally.

\subsection{ Central Production of glueballs}

In the double Pomeron exchange process each proton "emits" a Pomeron and the two Pomerons interact producing a massive system $M_X$. The massive system could form resonances or jet pairs. Because of the constraints provided by the double Pomeron interaction, glueballs, and other states coupling preferentially to gluons, could be produced with much reduced backgrounds compared to standard hadronic production processes.

In the kinematical region, which we are proposing to cover, those processes allow exploration of the non-perturbative regime of QCD. The strength of the STAR detector: excellent charged particle identification in the central rapidity region and $p_T$ resolution, coupled with ability to tag diffractive events with the forward protons with Roman pots. Central Production using Roman Pots and rapidity gap techniques has been studied at all the hadron colliders: ISR~\cite{Albrow86}, S$p\bar p$S~\cite{Brandt:2002qr} and the Tevatron~\cite{cdf} and is planned to be studied at the LHC~\cite{Albrow:2006xt}.

The idea that the production of glueballs is enhanced in the central region in the process $pp \to pM_{X}$p was first proposed by~\cite{close1} and was demonstrated experimentally~\cite{WA102}. The crucial argument here is that the pattern of resonances produced in central region, where both forward protons are tagged, depends on the vector difference of the transverse momentum of the final state protons $\vec k_{T1}$ and $\vec k_{T2}$, with $dP_T = |\vec k_{T1} - \vec k_{T2}|$. The so-called $dP_T$ filter argument is that when $dP_T$ is large ($\Lambda_{QCD}$) 
$q{\bar q}$ states are prominent and when $dP_T$ is small the surviving resonances include glueball candidates~\cite{close1,WA102}.

In what we are proposing large data samples of diffractive states can be obtained and analyzed as function of diffractive mass $M_X$ and $t$ ($d^2\sigma/dM_{X}^2dt$) for central production.

\section{Implementation Plan}

We will execute the above physics program in two phases. In both phases Roman Pots and STAR detector shall be used, fig.~\ref{layout}.
\subsection{Phase I}
The existing pp2pp experimental set-up, already installed at STAR, will measure spin dependence of both elastic scattering in an unexplored $t$ and $\xi$ range, with respect to what has been done already, and of Central Production described above, for which our studies found that there is good acceptance.

\subsubsection{Measurement of elastic scattering}

Using the capacity of existing power supplies one can run with optics of $\beta^*= 20 m$ and  at $\sqrt{s}=200 GeV$. This optics could extend the $t$ coverage to $0.003 < |t| < 0.03 (GeV/c)^2$. Reaching such a small $|t|$-value allows measuring the single spin analyzing power $A_N$ close to its maximum at $|t| \approx 0.0024 (GeV/c)^2$, where $A_{max} = 0.04(GeV/c)^2$, at $\sqrt{s}=200 GeV$. The $A_N$ and its $t$-dependence in the covered range is sensitive to a possible contribution of the single spin-flip amplitude, $\phi_5$~\cite{Trueman07}, from the interference between the hadronic spin-flip amplitude with the electromagnetic non-flip amplitude. An additional contribution of the hypothetical Odderon to the pp scattering amplitude can be probed by measuring the double spin-flip asymmetry, $A_{NN}$~\cite{Trueman07}. 

Given polarization 50$\%$ and 2.3 mb cross section within our acceptance we shall get $6.7\times10^6$ events. In the four $t$ subintervals we shall have $1.66 \times 10^6$ events in each. The corresponding errors are $\Delta A_N=0.0017, \Delta A_{NN}=\Delta A_{SS}=0.0053$. To estimate the error on $A_{NN}$ the $\phi$ intervals  $-45\deg <\phi < 45\deg$ and $135\deg <\phi < 225\deg$ were used, and complementary intervals for $A_{SS}$.

For the amount of data we expect to collect in 2008, an estimated error on the slope parameter is $\Delta b=0.31 (GeV/c)^{-2}$ and on the ratio of real to imaginary part $\Delta\rho=0.01$, which is comparable to the existing measurements from the $pp$ and $p \bar p$ data.

\subsubsection{Measurement of Central Production}
We studied the geometrical acceptance of our setup for both SDD and DPE processes. We have generated protons with $t$ and $\xi$ uniformly distributed in the regions $0.003 \leq |t| \leq 0.04 (GeV/c)^2$ and $0.005 \leq \xi \leq 0.05$ respectively. We assumed that the Roman Pots are 10mm from the beam center, which is at least $12\sigma$ of the beam size at the detection point. 

Our studies indicate that there is good acceptance to measure inelastic diffraction processes DPE with $\beta^* = 20m$ optics for Phase I. With the expected luminosity we can collect about $5 \times 10^6$, triggered DPE events. Number of events  for which the proton momentum is reconstructed, where it is required that two RPs on each side are used allowing reconstruction of the outgoing proton momentum, is about factor of ten lower. One assumes a 10 $\mu$barn cross section within our acceptance for the DPE process. 

Also, as noted earlier the events seen in the STAR Time Projection Chamber (TPC) in pp Central Production are very similar to those in heavy-ion UPC collisions. The algorithms and reconstruction code that has been developed to deal with those events can also be used in our program.  In particular experience gained in dealing with the backgrounds is very valuable.

\subsection{Phase II}

To maximize the acceptance and the range in $t$, $\xi$ and $M_X$ in this phase the Roman Pot system needs to be installed between DX-D0 magnets and will be used in conjunction with the STAR TPC to reconstruct and fully constrain events with resonance in central production process. A study needs to be done that range. The search for exotics is one of the topics of interest here, but not the only one.

\section{Summary}

In summary the physics program with tagged forward protons at STAR will:
1) Study elastic scattering and its spin dependence in unexplored $t$, $\xi$ and $\sqrt{s}$ range; 2)Study the structure of color singlet exchange in the non-perturbative regime of QCD; 3) Search for diffractive production of light and massive systems in double Pomeron exchange process; 4) Search for new physics, including glueballs and Odderon.

Finally we stress that the studies we are proposing will add to our understanding of QCD in the non-perturbative regime where calculations are not easy and one has to be guided by measurements.
\section{Acknowledgements}
The research reported here has been performed in part under the US DOE contract DE-AC02-98CH10886.
\end{document}